# Quantum transport of two-species Dirac fermions in dual-gated three-dimensional topological insulators

**Authors:** Yang Xu[1,2], Ireneusz Miotkowski[1], Yong P. Chen[1,2,3,]*

**Affiliations:**

[1]Department of Physics and Astronomy, Purdue University, West Lafayette, IN 47907 USA.

[2]Birck Nanotechnology Center, Purdue University, West Lafayette, IN 47907 USA.

[3]School of Electrical and Computer Engineering, Purdue University, West Lafayette, IN 47907 USA.

*Correspondence to: yongchen@purdue.edu

**Abstract:** Topological insulators are a novel class of quantum matter with a gapped insulating bulk yet gapless spin helical Dirac fermion conducting surface states. Here, we report local and non-local electrical and magneto transport measurements in dual-gated $BiSbTeSe_2$ thin film topological insulator devices, with conduction dominated by the spatially separated top and bottom surfaces, each hosting a single species of Dirac fermions with independent gate control over the carrier type and density. We observe many intriguing quantum transport phenomena in such a fully-tunable two-species topological Dirac gas, including a zero-magnetic-field minimum conductivity close to twice the conducatance quantum at the double Dirac point, a series of ambipolar two-component half-integer Dirac quantum Hall states and an electron-hole total filling factor zero state (with a zero-Hall plateau), exhibiting dissipationless (chiral) and dissipative (non-chiral) edge conduction respectively. Such a system paves the way to explore rich physics ranging from topological magnetoelectric effects to exciton condensation.

**Introduction**

A three-dimensional topological insulator (TI) is characterized by an insulating bulk band gap and gapless conducting topological surface states (TSS) of spin-helical massless 2D Dirac fermions[1,2]. Such surface states are topologically non-trivial and protected by time-reversal symmetry, thus immune to back scattering. The potential novel physics offered by this system, such as topological magnetoelectric (TME) effects[3,4], Majorana fermions[5], and effective magnetic monopoles[6], has drawn intense interest. One of the most iconic transport signatures for 2D Dirac electronic systems is the half-integer quantum Hall effect (QHE) in a perpendicular magnetic field ($B$), as first observed in graphene[7,8] and later also studied in HgTe[9,10]. The Landau Levels (LL) of 2D Dirac fermions have energies $E_N=\text{sgn}(N)v_F(2eB\hbar|N|)^{1/2}$, where sgn is the sign function, $N$ is the LL index (positive for electrons and negative for holes), $v_F$ is the Fermi velocity, $e$ is the elemental charge and $\hbar$ is the Plank's constant $h$ divided by $2\pi$. The zeroth LL at $E_0=0$ is equally shared between electrons and holes, giving rise to the half-integer shift in the quantized Hall conductivity $\sigma_{xy}=g(N+1/2)e^2/h$, where $g$ is the number of degenerate species of Dirac fermions (e.g., $g=4$ for graphene, and $g=1$ for TSS with a single Dirac cone). This 1/2 can also be related to the Berry-phase due to the spin or pseudospin locking to the momentum of Dirac fermions[7–10].

In most commonly studied TI materials such as $Bi_2Se_3$, $Bi_2Te_3$ and other Bi/Sb-based chalcogenides, it is often challenging to observe characteristic TSS transport (particularly QHE) due to bulk conduction caused by unintentional impurity doping. Only very recently has well-developed QHE arising from TSS been observed in exfoliated flakes from $BiSbTeSe_2$ (BSTS) single crystals[11] and molecular beam epitaxy grown $(Bi_{1-x}Sb_x)_2Te_3$ or $Bi_2Se_3$ thin films[12,13]. In this work, we fabricate dual-gated[14–16] TI devices from exfoliated BSTS thin flakes with undetectable bulk carrier density and conduction at low temperature[11]. Such a dual-gating structure is also promising for exploring exciton condensation proposed for TIs[17] and topological quantum phase transitions induced by displacement electric field[18].

In our dual-gated BSTS devices, the independent, ambipolar gating of parallel-conducting top and bottom surfaces realize two independently-controlled species of 2D Dirac fermions, allowing us to investigate such interesting transport phenomena as the minimum conductivity of TSS at Dirac point (DP), and two-species (two-component) Dirac fermion QHE of electron+electron, electron+hole and hole+hole types, involving various combinations of top and bottom surface half-integer filling factors $v_t$ and $v_b$. When $(v_t, v_b)=(-1/2, 1/2)$ or $(1/2, -1/2)$, there's an intriguing $v=0$ state characterized by zero Hall plateau and a large longitudinal resistance peak[11,12], attributed to the formation of dissipative and non-chiral edge states. We also perform non-local transport measurements and compare that with the normal local measurements in our dual-gated 3D TI devices in the quantum Hall regime to probe the nature of edge state transport for both standard QH states and the novel $v=0$ dissipative QH-like state. We further demonstrate that the dissipative edge states at $v=0$ have temperature-independent conductance, revealing that the transport in such a quasi-1D dissipative metallic edge channel could evade standard localization.

## Results

**Transport properties at zero and low magnetic field.** Qualitatively similar data are measured in multiple samples, while results from a typical sample A (channel length $L$=9.4 μm, width $W$=4.0 μm, with ~100 nm-thick BSTS and 40 nm-thick h-BN as top-gate dielectric, see schematic in Fig. 1a) are presented below unless otherwise noted. The h-BN as a substrate or gate dielectric is known to preserve good electronic properties for graphene, resulting from the atomic flatness and relatively low density of impurities in h-BN[19]. The carrier densities of the top and bottom surface of the BSTS flake are tuned by top-gate voltage $V_{tg}$ and back-gate voltage $V_{bg}$ respectively.

Fig. 1b and 1c show the double-gated electric field effect measured at $T$=0.3 K. The longitudinal resistivity $\rho_{xx}$ (=$R_{xx}*W/L$, with $R_{xx}$ being longitudinal resistance) at magnetic field $B$=0 T (Fig. 1b) and Hall resistivity $\rho_{xy}$ (=$R_{xy}$, Hall resistance) at $B$=1 T (Fig. 1c) are plotted in color scale as functions of both top and bottom gate voltages ($V_{tg}$ and $V_{bg}$). The extracted field effect and Hall mobilities are typically several thousands of cm$^2$V$^{-1}$s$^{-1}$. A minimum carrier density $n^*$ ~9×10$^{10}$ cm$^{-2}$ per surface can be extracted from the maximum Hall coefficient (absolute value) ~3.5 kΩ/T (when both surfaces are slightly n-type or p-type) measured in Fig. 1c. A set of exemplary $V_{tg}$-sweeps with $V_{bg}$=3 V is shown in Fig. 1c inset. By adjusting $V_{tg}$ (or $V_{bg}$), the device can be gated through a $R_{xx}$ peak, identified as the charge-neutrality DP of the top (or bottom) surface, marked by the blue (or red) dashed lines in Fig. 1b. Gating through the DP, the carriers in the corresponding surface change from hole-like to electron-like (i.e., ambipolar), as evidenced by Hall measurements (Fig. 1c). The slight deviation of the two lines from being perfectly vertical and horizontal arises from the weak capacitive coupling between the top (bottom) surface and the back (top) gate[16]. The crossing of these two lines corresponds to the double DP (both surfaces tuned to DP) where $\rho_{xx}$ ($\sigma_{xx}=1/\rho_{xx}$) reaches a global maximum (minimum). Within the gate voltage range used, the carriers predominantly come from the TSS and we observe relatively good particle-hole symmetry in the transport properties (e.g., the symmetrical appearance of $\rho_{xx}$ on both sides of DP in each surface in Fig. 1b and the similar absolute values of the positive and negative maximum Hall coefficient in Fig. 1c).

We have studied 6 dual-gated BSTS devices with different thicknesses ($t$) and aspect ratios ($L/W$). These devices are measured at low temperatures ($T<2$ K) and the results are repeatable after multiple thermal cycles. When both surfaces are tuned to DP, the minimum 2D conductivity $\sigma_{min}$ at $B=0$ T exhibits relatively constant value $(3.8\pm0.1)e^2/h$ for all the devices measured (with the uncertainty representing 90% confidence interval), whose thicknesses range from ~50 to ~200 nm and $L/W$ range from 1.3 to 3.5 (Fig. 1d). Our observation indicates that the conductivity at the DP for each major surface (top or bottom) is ~$2e^2/h$ (one unit of conductance quantum), within the range of values (2~5 $e^2/h$) reported by Kim et al.[14] on thin flakes of $Bi_2Se_3$ (~10 nm). The better consistency over multiple samples in our dual-gated BSTS devices may be attributed to the more insulating bulk (whose conduction is immeasurably small at low temperature) and uniformity of the exfoliated BSTS flakes, which are sandwiched between $SiO_2$ and h-BN to achieve better device stability. The minimum conductivity at DP has also been discussed in graphene with considerable interest[20–25].

The experiments in graphene revealed that the minimum conductivity is strongly affected by carrier-density inhomogeneities (puddles) induced by disorder on or near graphene[24,25], such as the absorbates or charged impurities in the substrates. In 3D TIs, one source of impurities likely relevant to the observed quasi-universal minimum conductivity in our dual-gated BSTS devices could be bulk defects (located near surface)[26,27], such as those revealed in scanning tunneling microscopy studies[28].

**Two-component quantum Hall effect.** For the rest of the paper, we focus on the transport phenomena in the quantum Hall (QH) regime under a high magnetic field $B$ perpendicular to the top and bottom surfaces. Fig. 2a and 2c show in color scales the longitudinal conductivity $\sigma_{xx}$ $(=\rho_{xx}/(\rho_{xx}^2+\rho_{xy}^2))$ and Hall conductivity $\sigma_{xy}$ $(=\rho_{xy}/(\rho_{xx}^2+\rho_{xy}^2))$ for Sample A as functions of $V_{tg}$ and $V_{bg}$ at $B$=18 T and $T$=0.3 K. The color plots in (**a**) and (**c**) divide the ($V_{tg}$, $V_{bg}$) plane into a series of approximate parallelograms, centered around well-developed or developing QH states with vanishing or minimal $\sigma_{xx}$ (**b**) and quantized $\sigma_{xy}$ in integer units of $e^2/h$ (**d**). These QH parallelograms are bounded by approximately (but slightly tilted) vertical and horizontal lines, which represent the top and bottom surface LLs respectively. By increasing (decreasing) either $V_{tg}$ or $V_{bg}$ to fill (exhaust) one LL on the top or bottom surface, $\sigma_{xy}$ increases (decreases) by $e^2/h$, taking consecutive quantized values of $ve^2/h$, where integer $v=v_t+v_b=N_t+N_b+1$. The $N_{t(b)}$ is the corresponding top (bottom) surface LL integer index that can be adjusted by top (back) gate to be of either Dirac electrons or holes. In Fig. 2d, different fixed $V_{bg}$ values (from -17 V to 40 V) set $v_b$ around consecutive half integers -3/2, -1/2, 1/2, 3/2 and 5/2 (such that the bottom surface contributes $\sigma_{xy}^b=v_b e^2/h$ to the total $\sigma_{xy}$), explaining the vertical shift of $e^2/h$ at QH plateaux of consecutive $V_{tg}$-sweeps.

It is also notable that in Fig. 2, there are a few states with zero quantized Hall conductivity ($\sigma_{xy}=0$, manifesting as white regions in Fig. 2c, separating the electron-dominated regions in red and the hole-dominated regions in blue) and non-zero $\sigma_{xx}$ minimum, marked by equal and opposite half-integer values of $v_t$ and $v_b$ thus total $v=0$, for example $(v_t, v_b)=(-1/2, 1/2)$, $(1/2, -1/2)$ and $(3/2, -3/2)$. These states with total $v=0$, exhibiting zero Hall plateaux (see also Fig. 2d), have non-zero $\sigma_{xx}$ minimum (Fig. 2a and 2b) but very large $R_{xx}$ maximum (see next, Fig. 3).

**Nonlocal transport at $v=0$ states.** To further characterize the observed QH and $v=0$ states, we have performed nonlocal transport measurements of $R_{nl}$ ($=V_{nl}/I$, $I$ is the current and $V_{nl}$ is the non-local voltage, see the schematic measurement setup in the inset of Fig. 3b) as functions of $V_{tg}$ and $V_{bg}$ at $B=18$ T and $T=0.3$ K and compared the results with the standard (local) measurements of the longitudinal resistance $R_{xx}$ (Fig. 3a). It's intriguing that unlike other QH states typified by a zero or minimum in $R_{xx}$, the states with $v=v_t+v_b=0$ (labeled by $(v_t, v_b)$ in Fig. 3a with $v_t=-v_b=\pm1/2$ or $\pm3/2$) are accompanied by a $R_{xx}$ maximum. The best-developed $v=0$ states are those at $(v_t, v_b)=(-1/2, 1/2)$ or $(1/2, -1/2)$, where $R_{xx}$ reaches ~220 kΩ ($\rho_{xx}$ ~100 kΩ), exceeding the resistance quantum ($h/e^2$=~25.8 kΩ) by an order of magnitude. The nonlocal $R_{nl}$ also becomes very large (~100 kΩ) and the similar order of magnitude as $R_{xx}$ at these two $v$=0 states while negligibly small at other ($v_t$, $v_b$) QH states (see Fig. 3b and also the representative cuts in Fig. 3c).

The simultaneously large local and non-local resistance at $v$=0 states in the QH regime has been reported in other 2D electron-hole systems[29,30] and understood in a picture of dissipative edge channels. We emphasize that the pronounced $R_{nl}$ signal cannot be explained from $R_{xx}$ by a classical Ohmic non-local resistance from the stray current connecting the remote leads. Such a contribution (=~$\rho_{xx}e^{-\pi L/W}$) would decay exponentially with $L/W$ (=2.4 in our case), and be three orders magnitude smaller than the local $R_{xx}$ (which is the case at $B$=0 T, Supplementary Fig. 1). As another comparison, the middle panel of Fig. 3c shows the cuts in Fig. 3(a,b) at $V_{bg}$=3 V, crossing the double-DP (also zeroth LL) of both top and bottom surfaces at ($v_t$, $v_b$)=(0, 0), where we observe a relatively large peak in $R_{xx}$ but significantly smaller $R_{nl}$. Such a result is consistent with the *extended* state transport (at the center of zeroth LL) as the current flows through the bulk of the 2D surface.

From the color plots in Figs. 2 and 3, the parallelogram centered around ($v_t$, $v_b$)=(-1/2, 1/2) state is enclosed by boundaries representing $N_t$=0 and -1, $N_b$=0 and 1 LLs. Similarly, ($v_t$, $v_b$)=(1/2, -1/2) state is bounded by $N_t$=0 and 1, $N_b$=0 and -1 LLs. We conclude that such a $v$=0 state can exist when the potential difference $V$ between top and bottom surfaces (equivalently the energy separation between top and bottom surface DPs) is in the range of $0<|V|<2E_{0-1}$ ($\approx 2\times 50$ meV at $B$=18 T, where $E_{0-1}$ is the 0-1 LL separation of TSS Dirac fermions[11]). The large energy scale of $E_{0-1}$ can help make the $v$=0 and $v$=±1 QH states observable at significantly elevated temperatures as demonstrated below.

**Temperature dependence of the $v$=0 and ±1 states.** We have studied the temperature ($T$) dependence of the QHE and $v$=0 states from 0.3 K to 50 K at $B$=18 T (Fig. 4). At each temperature, the bottom surface density is tuned by $V_{bg}$ to set $v_b$ near 1/2 (dashed lines) or -1/2 (solid lines), and the peaks in local $R_{xx}$ and nonlocal $R_{nl}$ corresponds to ($v_t$, $v_b$)=(-1/2, 1/2) or (1/2, -1/2), respectively (Fig. 4a and 4b, detailed raw data shown in Supplementary Fig. 2). The $R_{xx}$ peaks (>~150 kΩ) are seen to be more robust up to the highest temperature ($T$=50 K) measured while $R_{nl}$ peaks decrease rapidly (approximately linearly in $T$, shown in Fig. 4c) with increasing $T$ and is nearly suppressed above 50 K. We also show the $T$-dependence of $\sigma_{xx}$ and $\sigma_{xy}$ at ($v_t$, $v_b$)=(-1/2, 1/2), (1/2, -1/2), (1/2, 1/2) and (-1/2, -1/2) in Fig. 4e and 4f. The $\sigma_{xy}$ maintains good quantization at $ve^2/h$ ($v$=0, ±1) up to $T$=50 K while $\sigma_{xx}$ increases with $T$ (the gate-dependent $\sigma_{xx}$ and $\sigma_{xy}$ traces at different temperatures are shown in Supplementary Fig. 3). The $\sigma_{xx}$ for $v$=±1 states is found to show thermally activated behavior at high temperatures[11], where the finite $\sigma_{xx}$ is attributed to the thermal excited 2D surface or 3D bulk carriers. Such carriers can shunt the edge-state transport and suppress the nonlocal $R_{nl}$ response at high $T$[29]. We also note that the $\sigma_{xx}$ versus $T$ curves for $v$=0 and $v$=±1 states follow the similar trend and have approximately constant separation. We find the averaged separation $\Delta\sigma_{xx}$=1/2*[$\sigma_{xx}$(-1/2,1/2)+$\sigma_{xx}$(1/2,-1/2)-$\sigma_{xx}$(1/2,1/2)-$\sigma_{xx}$(-1/2,-1/2)] to be largely $T$-independent with a value of $(0.27\pm0.01)e^2/h$, which we attribute to the conductivity of the quasi-1D dissipative edge channel.

**Discussions**

In our measurement setup, the contacts connect to the top, bottom and side surfaces, all of which are probed simultaneously. The side surface only experiences an in-plane field and can be viewed as a quasi-1D domain boundary that separates the top and bottom surfaces with $B$ pointing outward and inward respectively, thus can support QH edge states[31]. When the top and bottom surfaces are doped to the same carrier type (either n or p), the corresponding QH edge states (on the side surface) would have the same chirality and give the observed total $\sigma_{xy}=ve^2/h=(v_t+v_b)e^2/h$, restricted to integer multiples of $e^2/h$. When the two surfaces have opposite carrier types but one of the them dominates, well-defined QH states with $v=v_t+v_b$ may still be observed, such as the (-1/2, 3/2) state with $\sigma_{xy}=(-1/2+3/2)e^2/h=e^2/h$ and vanishing $\rho_{xx}$. Previous studies in InAs/(AlSb)/GaSb heterostructure based electron-hole systems[32,33] also revealed QH effect with $R_{xy}=h/(ve^2)=h/(v_e-v_h)e^2$ ($v_e$ and $v_h$ are electron and hole filling factors, both are positive *integers*) and vanishing $R_{xx}$ when the AlSb barrier (separating electron and hole gases) is sufficiently thin to enable electron-hole hybridization. Despite the phenomenological similarities, our QH system is distinctive in the sense that the spatially separated electrons and holes residing on the top and bottom surfaces have *half* integer filling factors, and the hybridization only happens at the side surface.

We show the schematic energy spectrum when the two surfaces are degenerate with $V=0$[34–36] in Fig. 4g, which depicts the Fermi energy $E_f$ inside the 0-1 LL gap and corresponds to the (1/2, 1/2) QH state. For a relatively thick sample such as ours (~100 nm>>magnetic length $l_B=(\hbar/eB)^{1/2}\approx 6$ nm at $B=18$ T), however, it has been suggested that even in the presence of well quantized LLs, a standard TI Hall measurement would exhibit deviations from perfectly quantized values due to conduction through the side surfaces[31,34–36]. On the other hand, it has also been suggested that when net chiral modes exist (Fig. 4g and 4i show one such net chiral mode), the QH effect may be restored by the local equilibrium between non-chiral edge modes[37], possibly explaining the good quantization in $\sigma_{xy}$ and vanishing $\rho_{xx}$ (also $R_{nl}$) observed in our experiments.

For the $(v_t, v_b)= (-1/2, 1/2)$ or $(1/2, -1/2)$ state, the carrier density on the top and bottom surfaces are opposite. Since $E_f$ is within the LL gap on both the surfaces, the finite residual $\sigma_{xx}$ and large $R_{nl}$ we observed are indicative of dissipative edge transport. We show a schematic energy spectrum[38] of this $v=0$ state with $V$ slightly smaller than $E_{0-1}$ in Fig. 4h, where the Fermi level $E_f$ resides between the $N_t=-1$ and 0 LL of top surface (marked in blue), thus $v_t=-1/2$, and also between the $N_b=0$ and 1 LL of bottom surface, thus $v_b=1/2$. Overall, such energy spectrum represents a $(v_t, v_b)=(-1/2, 1/2)$ and $v=0$ state. The $E_f$ crosses an even number (only two shown in this illustrative example in Fig. 4h) of counter-propagating edge modes (arising from sub-bands of the quasi-1D side surface). The disorder can cause scattering and local equilibrium between the counter-propagating modes, giving rise to non-chiral dissipative transport (depicted by a series of conducting loops that can hop between adjacent ones in Fig. 4j) on the side surface with a large and finite resistance. While the energy spectrum (Fig. 4h) is expected to have a gap ($\Delta$) near the edge (due to the hybridization between top (marked with blue) and bottom (red) surface zeroth LLs and approximately the finite-size confinement-induced gap $\approx hv_F/t \approx 10$ meV opened at DP of the side surface[38]), we did not observe a truly insulating state with vanishing $\sigma_{xx}$ and diverging $R_{xx}$ (Figs. 2a, 4a). This is likely due to the disorder potential (spatial fluctuation of DP[28]) comparable or larger than $\Delta$ and thus smearing out this gap (effectively $E_f$ always crosses the non-chiral edge modes). It would be an interesting question for future studies to clarify whether the weak $T$-dependence (at $T<\sim 50$ K) of the observed conductance (Fig. 4e), similar to the behavior reported in InAs/GaSb based electron-hole systems[39], may indicate an absence of localization[40–43] in such quasi-1D resistive edge channels.

Several recent theories have pointed out that the $v=1/2-1/2=0$ state in the TI QH system may bring unique opportunities to realize various novel physics. It has been suggested that both the $v=0$ state in TI QHE and an analogous quantum anomalous Hall (QAH) state with zero-Hall-conductance plateau in a magnetic doped TI around the coercive field can be used as platforms to observe the TME effect[44,45], where an electric (magnetic) field induces a co-linear magnetic (electric) polarization with a quantized magnetoelectric polarizability of $\pm e^2/2h$. A zero-Hall-plateau state has been recently observed in the QAH case in ultrathin (few-nm-thick) films of $Cr_x(Bi,Sb)_{2-x}Te_3$ at low temperature ($<1$ K)[46,47]. In comparison, our samples have much larger thickness ($>\sim 50$ nm, suggested to be preferable for better developed TME effect[45,48]), and our $v=0$ state survives at much higher temperatures ($\sim 50$ K). It has also been proposed that excitonic condensation and superfluidity can occur in thin 3D TIs at the $v=0$ state in QH regime[49] (in addition to at zero $B$ field[17]) induced by spontaneous coherence between strongly-interacting top and bottom surfaces. In future studies, much thinner samples are likely needed to investigate the possibility of such exciton superfluidity.

**Methods**

**Sample preparation.** Three-dimensional topological insulator (3D TI) single crystals BiSbTeSe$_2$ (BSTS) were grown by the vertical Bridgman technique[11]. BSTS flakes (typical thickness ~50-200 nm) are exfoliated ("Scotch tape method") onto highly doped Si (p+) substrates (with 300 nm-thick SiO$_2$ coating), and lithographically fabricated into Hall bar shaped devices with Cr/Au contacts. A thin flake of hexagonal boron nitride (h-BN, typical thickness ~10-40 nm) is transferred[19] on the BSTS flake to serve as a top-gate dielectric and a top-gate metal (Cr/Au) is deposited afterwards. The thickness of BSTS and h-BN flakes are measured by atomic force microscopy.

**Transport measurement.** Transport measurements are performed with the standard lock-in technique using a low-frequency (<20 Hz) excitation current of 20 nA in a helium-4 variable temperature system (with base temperature down to 1.6 K) or a helium-3 system equipped with magnetic fields ($B$) up to 18 T (down to 0.3 K).

**References**

1.  Hasan, M. Z. & Kane, C. L. Colloquium: Topological insulators. *Rev. Mod. Phys.* **82,** 3045–3067 (2010).
2.  Qi, X.-L. & Zhang, S.-C. Topological insulators and superconductors. *Rev. Mod. Phys.* **83,** 1057–1110 (2011).

3. Qi, X. L., Hughes, T. L. & Zhang, S. C. Topological field theory of time-reversal invariant insulators. *Phys. Rev. B* **78,** 195424 (2008).

4. Essin, A. M., Moore, J. E. & Vanderbilt, D. Magnetoelectric polarizability and axion electrodynamics in crystalline insulators. *Phys. Rev. Lett.* **102,** 146805 (2009).

5. Fu, L. & Kane, C. L. Superconducting Proximity Effect and Majorana Fermions at the Surface of a Topological Insulator. *Phys. Rev. Lett.* **100,** 096407 (2008).

6. Qi, X.-L., Li, R., Zang, J. & Zhang, S.-C. Inducing a magnetic monopole with topological surface States. *Science* **323,** 1184–1187 (2009).

7. Novoselov, K. S. *et al.* Two-dimensional gas of massless Dirac fermions in graphene. *Nature* **438,** 197–200 (2005).

8. Zhang, Y., Tan, Y.-W., Stormer, H. L. & Kim, P. Experimental observation of the quantum Hall effect and Berry's phase in graphene. *Nature* **438,** 201–204 (2005).

9. Büttner, B. *et al.* Single valley Dirac fermions in zero-gap HgTe quantum wells. *Nat. Phys.* **7,** 418–422 (2011).

10. Brüne, C. *et al.* Quantum Hall Effect from the Topological Surface States of Strained Bulk HgTe. *Phys. Rev. Lett.* **106,** 126803 (2011).

11. Xu, Y. *et al.* Observation of topological surface state quantum Hall effect in an intrinsic three-dimensional topological insulator. *Nat. Phys.* **10,** 956–963 (2014).

12. Yoshimi, A. R. *et al.* Quantum Hall Effect on Top and Bottom Surface States of Topological Insulator Films. *Nat. Commun.* **6,** 6627 (2014).

13. Koirala, N. *et al.* Record Surface State Mobility and Quantum Hall Effect in Topological Insulator Thin Films via Interface Engineering. *Nano Lett.* **15,** 8245–8249 (2015).

14. Kim, D. *et al.* Surface conduction of topological Dirac electrons in bulk insulating $Bi_2Se_3$. *Nat. Phys.* **8,** 460–464 (2012).

15. Chang, C.-Z. *et al.* Simultaneous Electrical-Field-Effect Modulation of Both Top and Bottom Dirac Surface States of Epitaxial Thin Films of Three-Dimensional Topological Insulators. *Nano Lett.* **15,** 1090–1094 (2015).

16. Fatemi, V. *et al.* Electrostatic Coupling between Two Surfaces of a Topological Insulator Nanodevice. *Phys. Rev. Lett.* **113,** 206801 (2014).

17. Seradjeh, B., Moore, J. E. & Franz, M. Exciton Condensation and Charge Fractionalization in a Topological Insulator Film. *Phys. Rev. Lett.* **103,** 066402 (2009).

18. Kim, M., Kim, C. H., Kim, H.-S. & Ihm, J. Topological quantum phase transitions driven by external electric fields in $Sb_2Te_3$ thin films. *Proc. Natl. Acad. Sci.* **109,** 671–674 (2012).

19. Dean, C. R. *et al.* Boron nitride substrates for high-quality graphene electronics. *Nat. Nanotechnol.* **5,** 722–726 (2010).

20. Das Sarma, S., Adam, S., Hwang, E. H. & Rossi, E. Electronic transport in two-dimensional graphene. *Rev. Mod. Phys.* **83,** 407–470 (2011).

21. Nilsson, J., Neto, A. H. C., Guinea, F. & Peres, N. M. R. Electronic Properties of Graphene Multilayers. *Phys. Rev. Lett.* **97,** 266801 (2006).

22. Miao, F. *et al.* Phase-Coherent Transport in Graphene Quantum Billiards. *Science* **317 ,** 1530–1533 (2007).

23. Tan, Y.-W. *et al.* Measurement of Scattering Rate and Minimum Conductivity in Graphene. *Phys. Rev. Lett.* **99,** 246803 (2007).

24. Martin, J. *et al.* Observation of electron-hole puddles in graphene using a scanning single-electron transistor. *Nat. Phys.* **4,** 144–148 (2008).

25. Chen, J.-H. *et al.* Charged-impurity scattering in graphene. *Nat. Phys.* **4,** 377–381 (2008).

26. Li, Q., Rossi, E. & Das Sarma, S. Two-dimensional electronic transport on the surface of three-dimensional topological insulators. *Phys. Rev. B* **86,** 235443 (2012).

27. Skinner, B., Chen, T. & Shklovskii, B. I. Effects of bulk charged impurities on the bulk and surface transport in three-dimensional topological insulators. *J. Exp. Theor. Phys.* **117,** 579–592 (2013).

28. Beidenkopf, H. *et al.* Spatial fluctuations of helical Dirac fermions on the surface of topological insulators. *Nat. Phys.* **7,** 939–943 (2011).

29. Gusev, G. M. *et al.* Nonlocal Transport Near Charge Neutrality Point in a Two-Dimensional Electron-Hole System. *Phys. Rev. Lett.* **108,** 226804 (2012).

30. Nichele, F. *et al.* Insulating State and Giant Nonlocal Response in an InAs/GaSb Quantum Well in the Quantum Hall Regime. *Phys. Rev. Lett.* **112,** 036802 (2014).

31. Chu, R.-L., Shi, J. & Shen, S.-Q. Surface edge state and half-quantized Hall conductance in topological insulators. *Phys. Rev. B* **84,** 085312 (2011).

32. Mendez, E. E., Esaki, L. & Chang, L. L. Quantum Hall Effect in a Two-Dimensional Electron-Hole Gas. *Phys. Rev. Lett.* **55,** 2216–2219 (1985).

33. Suzuki, K., Miyashita, S. & Hirayama, Y. Transport properties in asymmetric InAs/AlSb/GaSb electron-hole hybridized systems. *Phys. Rev. B* **67,** 195319 (2003).

34. Lee, D.-H. Surface States of Topological Insulators: The Dirac Fermion in Curved Two-Dimensional Spaces. *Phys. Rev. Lett.* **103,** 196804 (2009).

35. Vafek, O. Quantum Hall effect in a singly and doubly connected three-dimensional topological insulator. *Phys. Rev. B* **84,** 245417 (2011).

36. Zhang, Y.-Y., Wang, X.-R. & Xie, X. C. Three-dimensional topological insulator in a magnetic field: chiral side surface states and quantized Hall conductance. *J. Phys. Condens. Matter* **24,** 015004 (2012).

37. Brey, L. & Fertig, H. A. Electronic states of wires and slabs of topological insulators: Quantum Hall effects and edge transport. *Phys. Rev. B* **89,** 085305 (2014).

38. Morimoto, T., Furusaki, A. & Nagaosa, N. Charge and Spin Transport in Edge Channels of a $\nu = 0$ Quantum Hall System on the Surface of Topological Insulators. *Phys. Rev. Lett.* **114,** 146803 (2015).

39. Takashina, K. *et al.* Insulating states of a broken-gap two-dimensional electron-hole system. *Phys. Rev. B* **68,** 235303 (2003).

40. Anderson, P. W. Absence of Diffusion in Certain Random Lattices. *Phys. Rev.* **109,** 1492–1505 (1958).

41. Kurkijärvi, J. Hopping conductivity in one dimension. *Phys. Rev. B* **8,** 922–924 (1973).

42. Lee, P. A. Variable-range hopping in finite one-dimensional wires. *Phys. Rev. Lett.* **53,** 2042–2045 (1984).

43. Dunlap, D. H., Kundu, K. & Phillips, P. Absence of localization in certain statically disordered lattices in any spatial dimension. *Phys. Rev. B* **40,** 10999–11006 (1989).

44. Morimoto, T., Furusaki, A. & Nagaosa, N. Topological magnetoelectric effects in thin films of topological insulators. *Phys. Rev. B* **92,** 085113 (2015).

45. Wang, J., Lian, B., Qi, X.-L. & Zhang, S.-C. Quantized topological magnetoelectric effect of the zero-plateau quantum anomalous Hall state. *Phys. Rev. B* **92,** 081107 (2015).

46. Feng, Y. *et al.* Observation of the Zero Hall Plateau in a Quantum Anomalous Hall Insulator. *Phys. Rev. Lett.* **115,** 126801 (2015).

47. Kou, X. *et al.* Metal-to-insulator switching in quantum anomalous Hall states. *Nat. Commun.* **6,** 8474 (2015).

48. Baasanjav, D., Tretiakov, O. A. & Nomura, K. Magnetoelectric effect in topological insulator films beyond the linear response regime. *Phys. Rev. B* **90,** 045149 (2014).

49. Tilahun, D., Lee, B., Hankiewicz, E. M. & MacDonald, A. H. Quantum Hall Superfluids in Topological Insulator Thin Films. *Phys. Rev. Lett.* **107,** 246401 (2011).

## Acknowledgements

We thank J. Hu, T. Wu, E. Palm, T. Murphy, A. Suslov, E. Choi and B. Pullum for experimental assistance. We also thank F. de Juan, R. Ilan, N. Nagaosa and W. Ku for helpful discussions. This work is supported by the DARPA MESO program (Grant N66001-11-1-4107). A portion of this work was performed at the National High Magnetic Field Laboratory, which is supported by National Science Foundation Cooperative Agreement No. DMR-1157490, the State of Florida, and the U.S. Department of Energy.

## Author contributions

Y.P.C supervised the research. I.M. synthesized the crystals. Y.X. fabricated the devices, performed the transport measurements and analyzed the data. Y.X. and Y.P.C wrote the paper.

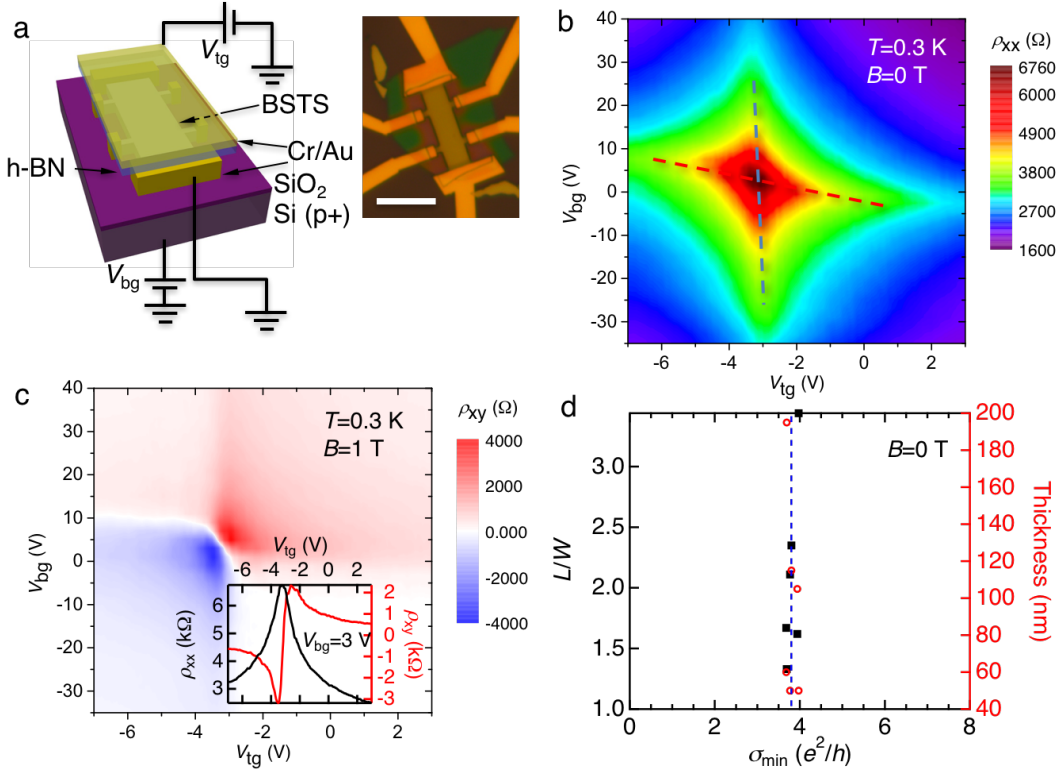

**Figure 1 | Device configuration and dual-gated field effect at zero and low magnetic field.** (**a**) Device schematic. Inset is an optical microscope image of a typical dual-gated BSTS device before depositing the top gate metal with scale bar 10 μm. (**b**, **c**) show 2D maps of $\rho_{xx}$ at $B=0$ T and $\rho_{xy}$ at $B=1$ T as functions of $V_{tg}$ and $V_{bg}$ on sample A. The blue (red) dashed lines in (**b**) are guides to the eye for the top (bottom) surface DP. The 2D map is generated by data measured from $V_{tg}$ sweeps at a series of $V_{bg}$ values, with one example at $V_{bg}=3$ V shown in the inset of (**c**). (**d**) Zero-magnetic-field minimum conductivity $\sigma_{min}$ (bottom axis) measured in 6 dual-gated samples at low temperature (<2 K) plotted as a function of the sample thickness (data in circles) and 2D aspect ratio ($L/W$, data in squares). The vertical dashed line indicates $3.8e^2/h$.

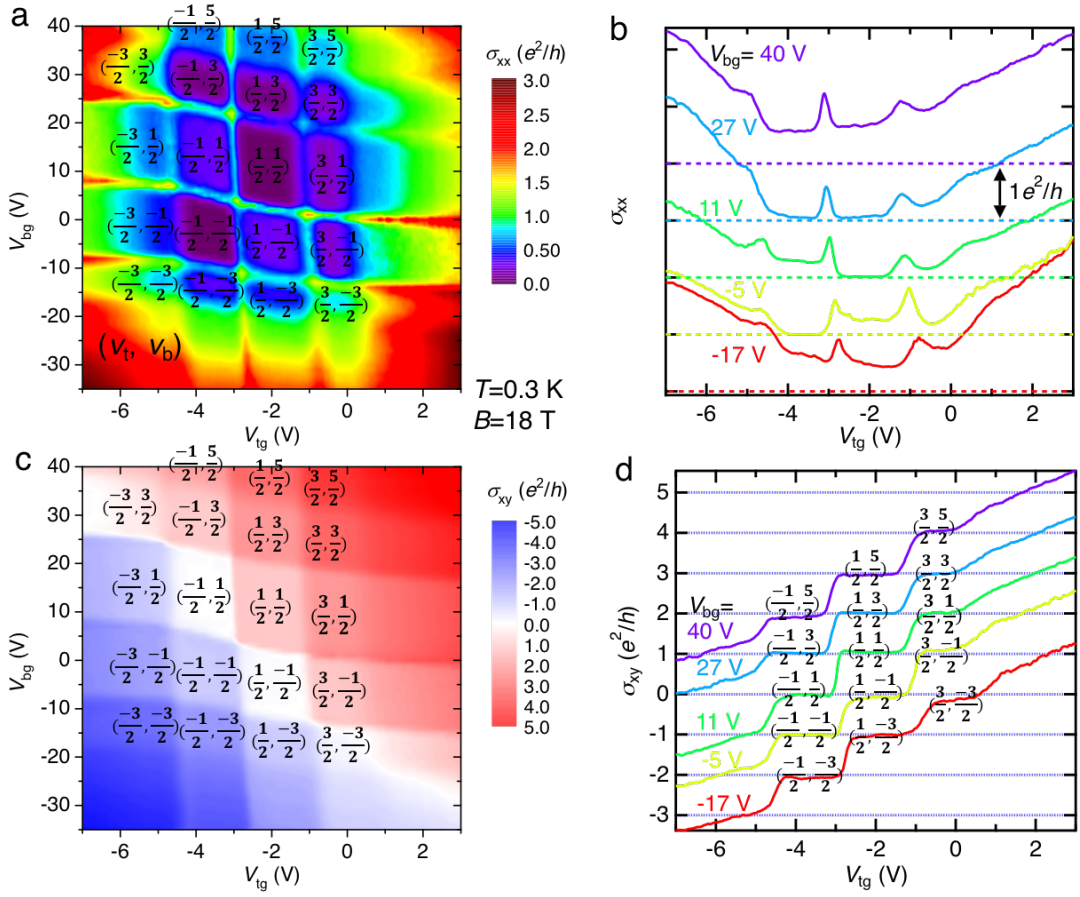

**Figure 2 | Quantum Hall effect modulated by top and bottom gates.** (**a**) $\sigma_{xx}$ and (**c**) $\sigma_{xy}$, shown as 2D color maps, as functions of $V_{tg}$ and $V_{bg}$ at $B$=18 T and $T$=0.3 K in sample A, with representative cuts at 5 different values of $V_{bg}$ shown in (**b**) and (**d**). The ($v_t$, $v_b$) labels (top, bottom) surface filling factors for corresponding quantum Hall states. The $\sigma_{xx}$ curves in (**b**) are shifted vertically (in consecutive step of $e^2/h$) for clarity (the corresponding zero $\sigma_{xx}$ levels are indicated by the same-colored horizontal dashed lines).

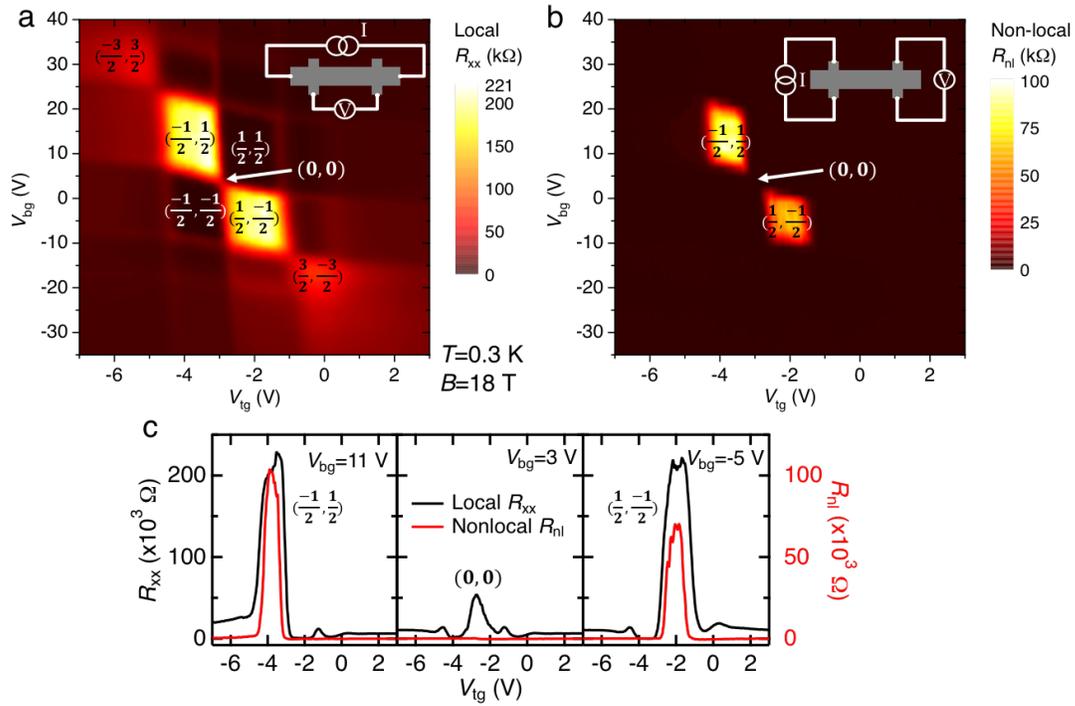

**Figure 3 | Local and non-local resistance in dual-gated TI in high magnetic field.** (**a**) Local resistance $R_{xx}$ and (**b**) non-local resistance $R_{nl}$ measured in sample A as functions of $V_{tg}$ and $V_{bg}$ at $B$=18 T and $T$=0.3 K, with insets showing the measurement setup schematics. (**c**) A few representative cuts of (**a**) and (**b**) at different values of $V_{bg}$. Filling factors for the local $R_{xx}$ peaks are labeled in each sub-panel.

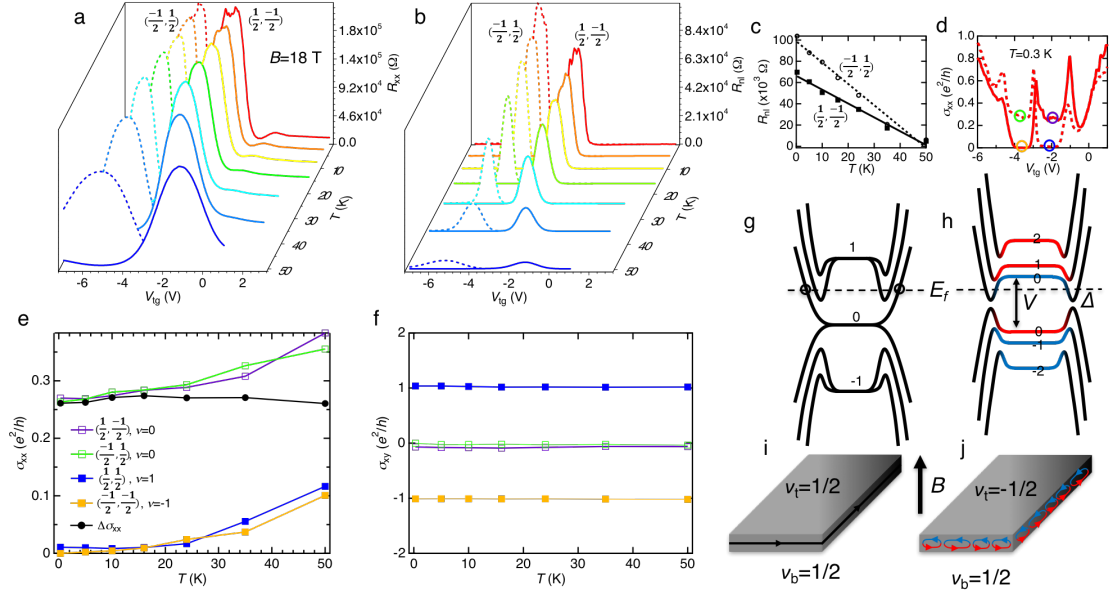

**Figure 4 | Temperature dependence and illustrative schematics of the QHE and $v$=0 state in TI.** (**a**) $R_{xx}$ and (**b**) $R_{nl}$ measured in sample A as functions of $V_{tg}$ for different temperatures at $B$=18 T, where $V_{bg}$ is chosen to set $v_b$ at 1/2 (dashed lines) and -1/2 (solid lines), respectively. (**c**) The $R_{nl}$ value at $(v_t, v_b)$=(-1/2, 1/2) and (1/2, -1/2) shows approximately linear dependence on temperature. (**d**) $\sigma_{xx}$ versus $V_{tg}$ (with the same two values of $V_{bg}$ chosen in (a) and (b)) at $T$=0.3 K as an example, with each highlighted circle corresponding to a state in (**e**) plotted with corresponding colored symbols. (**e**) $\sigma_{xx}$ and (**f**) $\sigma_{xy}$ of $v$=+1, -1 and 0 states as functions of temperature. In (**e**), we also plot $\Delta\sigma_{xx}$ (difference between averaged $v$=0 states' $\sigma_{xx}$ and averaged $v$=±1 states' $\sigma_{xx}$), which barely changes with $T$. (**g**,**h**) Schematics of surface band structure (energy spectrum) in high magnetic field, showing Landau levels from top and bottom surfaces (blue and red) in the middle of the sample transitioning into side surface sub-bands at sample edge, and (**i**,**j**) edge states in a slab-shaped sample for $v$=1 and $v$=0 states. The dashed line indicates a representative Fermi level $E_f$ and circles in (**g**) label chiral edge modes.

# Supplementary Information

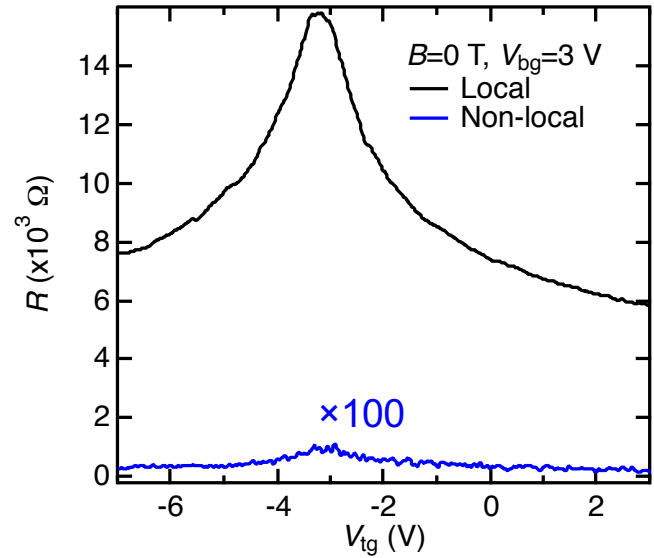

**Supplementary Figure 1 | Measured local and nonlocal resistance without magnetic field.** Local resistance $R_{xx}$ (a) and non-local resistance $R_{nl}$ (b) as functions of $V_{tg}$ at $V_{bg}$=3 V and $B$=0 T, $T$=0.3 K. The non-local resistance $R_{nl}$ shown here is multiplied by a factor of hundred. The data shown in this supplementary file are all from the same Sample A as that in the main text.

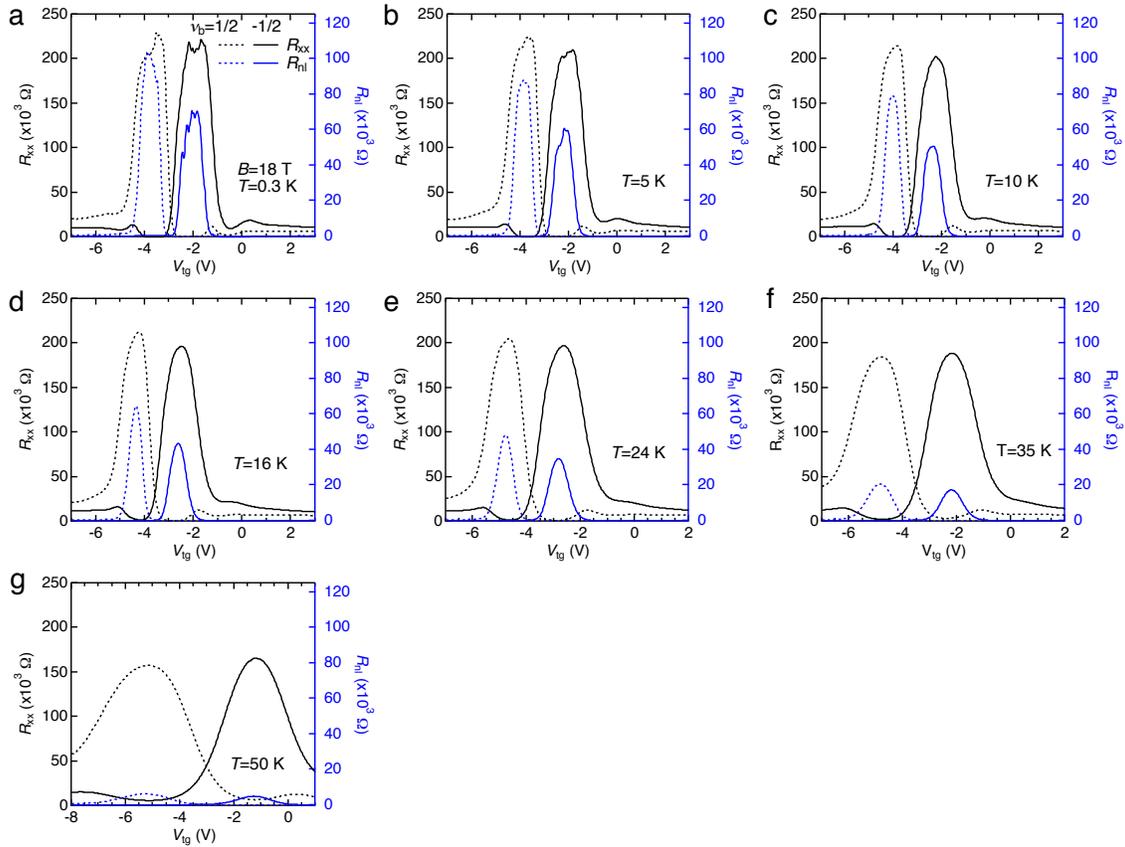

**Supplementary Figure 2 | Temperature dependence of the local and nonlocal resistance.** (a)-(g) Local resistance $R_{xx}$ and nonlocal resistance $R_{nl}$ versus $V_{tg}$ at fixed bottom surface filling factor $v_b$=1/2 (dashed) and -1/2 (solid) measured at $B$=18 T at different temperatures. At total filling factor $v$=0 states, the nonlocal resistance $R_{nl}$ decreases more rapidly with increasing temperature while the local resistance $R_{xx}$ maintains large value up to 50 K.

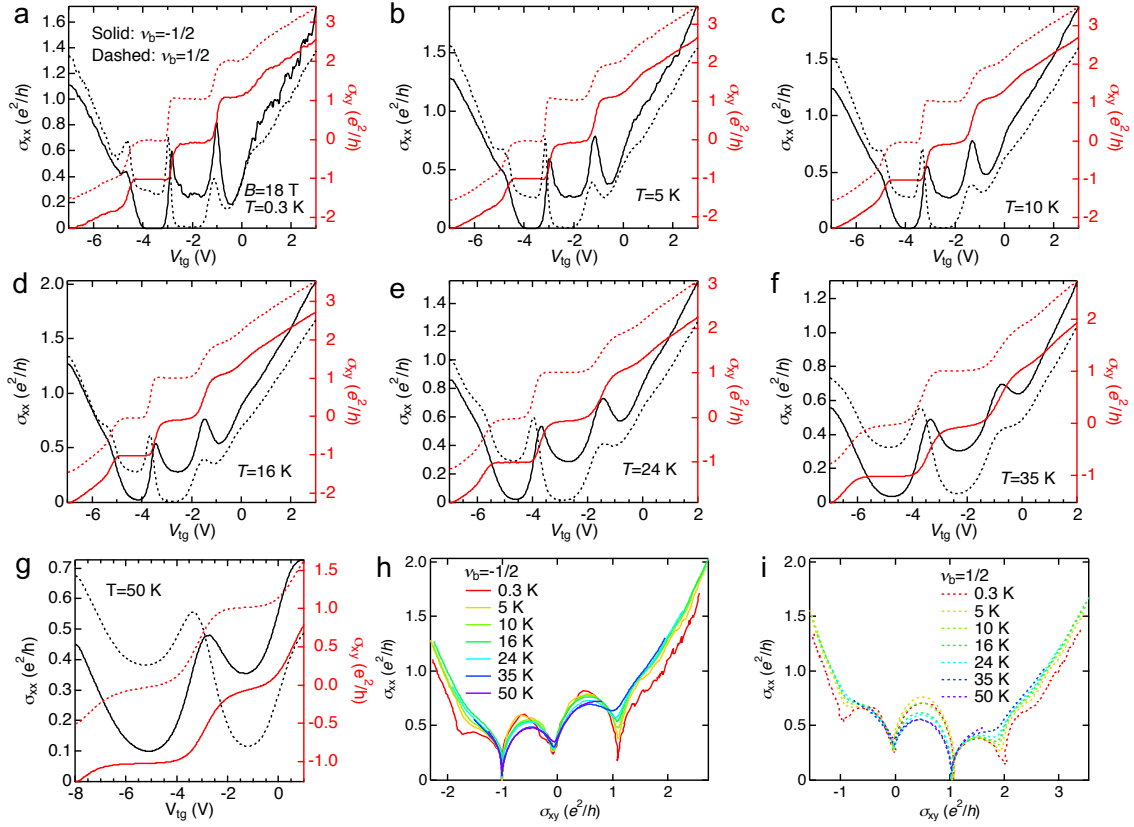

**Supplementary Figure 3 | Temperature dependence of the longitudinal and Hall resistivity.** (a)-(g) Longitudinal conductivity $\sigma_{xx}$ and Hall conductivity $\sigma_{xy}$ versus $V_{tg}$ at fixed bottom surface filling factor $\nu_b=1/2$ (dashed) and -1/2 (solid) measured at $B=18$ T at different temperatures. (h) and (i) show $\sigma_{xx}$ versus $\sigma_{xy}$ at $\nu_b=-1/2$ and $1/2$ respectively at different temperatures.